\documentclass[12pt]{article}

\usepackage{amsmath}
\usepackage{amssymb}
\usepackage{graphicx}
\usepackage{bm}
\usepackage{upgreek}
\begin{document}

\begin{center}

{\noindent \bf \large
Temperature variation in the dark cosmic fluid in the late universe }

\bigskip

Iver Brevik\footnote{iver.h.brevik@ntnu.no}

\bigskip
Department of Energy and Process Engineering, Norwegian University of Science and Technology, N-7491 Trondheim, Norway

\bigskip

\today

\end{center}

\bigskip

\begin{abstract}
 A one-component dark energy fluid model of the late universe is considered ($w<-1$) when the fluid, initially assumed laminar, makes a transition into a turbulent state of motion. Spatial isotropy is assumed so that only the bulk viscosities are included ($\zeta$ in the laminar epoch and $\zeta_{\rm turb}$ in the turbulent epoch). Both viscosities are assumed to be constants.  We derive a formula, new as far as we know, for the time dependence of  the temperature $T(t)$ in the laminar case when viscosity is included. Assuming that  the laminar/turbulent transition takes place at some time $t_s$ before the big rip is reached, we then analyze the positive temperature jump experienced by the fluid at $t=t_*$ if $\zeta_{\rm turb} > \zeta$. This is just as one would expect physically. The corresponding  entropy production is also considered. A special point emphasized in the paper is the analogy that exists between the cosmic fluid and a so-called Maxwell fluid in viscoelasticity.
 \end{abstract}



\section{	Introduction} \label{sec:1}

We begin by making some remarks on the theory of a {\it viscoelastic fluid} in hydromechanics. The reason for this rather uncommon approach  to a study in cosmology is that this kind of fluid bears some formal  similarity with the kind of cosmic fluids one usually is confronted with.  Viscoelastic fluids have in general a rich structure. The Navier-Stokes equation   for such a fluid in nonrelativistic approximation can be written in the form
\begin{equation}
\frac{\partial \bf u}{\partial t}+({\bf u\cdot \nabla)u}=-\frac{1}{\rho}\nabla p+\nu^* \nabla^2\bf u, \label{1}
\end{equation}
where $\nu^*$ is the generalized kinematic viscosity, given on operator form as
\begin{equation}
\nu^*=\nu\frac{1+\mu\,\partial/\partial t}{1+\lambda \,\partial/\partial t}, \label{2}
\end{equation}
$\nu=\eta/\rho$  being the  ordinary (molecular) kinematic viscosity and   $\eta$  the laminar  shear viscosity. The coefficients $\mu$ and $\lambda$ characterize the viscoelastic properties. When $\lambda>0, \mu=0$ the fluid is called a  Maxwell fluid;  when $\lambda>0, \mu>0$ it is called an   Oldroyd fluid.  The expression (\ref{2}) refers to first order viscoelasticity only; if higher  order viscoelastic fluids were considered, higher order derivatives must be included.
The theory of viscoelastic fluids can be found in various references; we here follow the conventions of Ref.~\cite{das02}.

We will henceforth consider the simplest case,  a Maxwell fluid, only.  Writing for simplicity the left hand side of Eq.~(\ref{1}) as $d{\bf u}/dt$ with $d/dt$ meaning the total derivative, we then have
\begin{equation}
\left(1+\lambda \frac{\partial}{\partial t}\right)\frac{d\bf u}{dt}=-\frac{1}{\rho}\left(1+\lambda \frac{\partial}{\partial t}\right)\nabla p+\nu \nabla^2\bf u. \label{3}
\end{equation}
A noteworthy property of this equation is that it is well suited to describe sudden transitions that may take case,  for instance, in connection with phase transitions.  Let such a  transition take place at a definite instant $t=t_*$. Integrating Eq.~(\ref{3}) over time from $t_*^-$ to $t_*^+$, we see that the presence of $\lambda$ permits the left hand side to describe a jump in the acceleration $\partial {\bf u}/\partial t$ (the velocity $\bf u$ itself does not jump). Similarly, the first term on the right hand side permits description of the jump in the pressure gradient as well as the pressure itself. The last term on the right is expected to be of minor importance.

  One of the objectives of the present paper is to point out that the viscoelastic property shown in Eq.~(\ref{3}) can be made use of in the phenomenological theory of the late universe, extending from present time $t=0$ into the future. Especially we will focus on the expected transition to a state of {\it turbulent flow} at some definite, though unspecified instant, $t=t_*$, identified to be the same as  the viscoelastic transition point $t_*$ above.

This outline calls for some background comments. First, one may ask: why consider the late universe at all, since astrophysical observations necessarily look back in time? A main reason for the recent interest in this topic comes from the observations of the equation-of-state parameter $w$ in the universe. We will assume, as usual, that the equation of state is homogeneous, of the form
\begin{equation}
p=w\rho, \quad w=\rm constant. \label{4}
\end{equation}
According to the 2015 Planck data \cite{planck} (Table 5) one has
\begin{equation}
w=-1.019^{+0.075}_{-0.080}. \label{5}
\end{equation}
This means that it is quite possible that the cosmic fluid is a phantom fluid, characterized by a value of $w$ that is less than $-1$. Let us in the following assume that the  fluid is a one-component dark energy fluid; writing
\begin{equation}
w=-1+\alpha, \label{6}
\end{equation}
we  see that $|\alpha|$ is small, of order $10^{-1}$ typically, though its sign is uncertain.
For some years, from the discovery of Caldwell et al. \cite{caldwell03},  it has been known that if the cosmic fluid starts out from a value of $w$ lying in the phantom region, it will encounter some form of singularity in the remote future. The most dramatic event is called the big rip, in which the fluid enters into a singularity after a finite time span \cite{caldwell03,nojiri03,nojiri04}. There are also softer variants of the future singularity where the singularity is not reached until an infinite time, called the little rip \cite{frampton11,brevik11,brevik12},  the pseudo rip \cite{frampton12}, and the quasi rip \cite{wei12}. Theories of this sort are of course speculative, but they are  founded on observations nevertheless.

The second ingredient in our theory, as mentioned,  is the transition to a turbulent state of motion in the late universe at some  instant $t_*$.  From a hydrodynamical viewpoint  such an assumption is physically natural, as violent motions and large densities are expected as the singularity is reached.  We have considered turbulence aspects in cosmology in earlier works; cf. Refs. \cite{brevik13,brevik12a,brevik11a}. Making use of the viscoelastic analogy, we will have the opportunity to calculate the entropy production at the transition.

What can be said about determining the onset of turbulence? The characteristic feature of turbulent flow is that it is unstable with respect to infinitesimal disturbances. The theoretical problem of the stability of steady flow has actually not been solved in general. In usual fluid dynamics turbulence occurs when the Reynolds number  Re  exceeds an experimentally determined critical value, for instance Re$_{\rm crit}=2300$ for pipe flow. An analogous  determination of the critical Reynolds number is not possible in the present case, both because the lack of an external geometric scale  and also because of  lack of knowledge about the microstructure (i.e., the particle motion, or eddy size) of the fluid. A macroscopic theory of the  fluid is unable to determine the transition point. A promising  way to improvement  might  here  be to include the particle motion explicitly in the comoving frame, in the way pointed out   by Bini et al. \cite{bini13}. If knowledge about  the local motion is achieved in this way, one  has  in principle the opportunity to reintroduce the Reynolds number concept by using the typical size of eddies as the length scale. We will however  leave that point here without further consideration, and  will in the following simply assume  that the transition point $t=t_*$ is determined phenomenologically, as a definite though unspecified point.

In order to make predictions about the late universe, one needs  to draw into consideration available information  about  the bulk viscosity at the present time $t=0$. The recent analysis of Wang and Meng \cite{wang14} is useful in this context, as the authors compare the theoretical curve for the Hubble parameter  $H=H(z)$    as   function of the redshift $z$  with a number of observations (actually inserting various forms for  a time-dependent $\zeta (t)$). As a mean extracted from their analysis, we  take
\begin{equation}
\zeta_0 \sim 10^5~{\rm Pa~ s} \label{7}
\end{equation}
to be a reasonable value at the present time (subscript zero refers to $t=0$).  This is also roughly in agreement with the analysis of Velten and Schwarz on dark matter dissipation \cite{velten12}. These authors   draw into consideration recent data from supernovae, baryon acoustic  oscillations, and cosmic microwave background, and  concludes  that dark matter has  a bulk viscosity less than about $10^7~$Pa s. An noteworthy point in this connection is that  they find  viscous theory  to permit galactic halos under the condition $24\pi G\zeta_0/{H_0}
  \ll 0.2$,     or  $\zeta_0 \ll 10^6~$Pa s.

   Another recent and valuable contribution to the literature is the recent paper of Sasidharan and Mathew \cite{sasidharan15}. They perform a phase space analysis of the universe with bulk viscosity of the form (i) $\zeta=\zeta_0$ a constant, (ii) $\zeta_0+\zeta_1\dot{a}/a$, (iii) $\zeta=\zeta_0+\zeta_1\dot{a}/a+\zeta_2\ddot{a}/\dot{a}$. Of interest here is option (i), for which we extract from table 1 in their paper
   \begin{equation}
   \frac{24\pi G}{c^2}\frac{\zeta_0}{H_0}=1.92,
   \end{equation}
   here written in dimensional units on the left hand side. Thus $\zeta_0$ becomes in this analysis quite large,
   \begin{equation}
   \zeta_0=7.57\times 10^7~\rm {Pa~s}.
   \end{equation}
   After all, as a conservative estimate, we suggest for the universe that the viscosity lies within a quite wide interval
   \begin{equation}
10^4~{\rm Pa~ s}<\zeta_0 < 5\times 10^7~{\rm Pa~ s},
\end{equation}
roughly in agreement also with our recent analysis \cite{brevik15}. This is of course many orders of magnitude higher than the bulk viscosity for water at atmospheric pressure and room temperature. We may also mention that  the best-fit analysis of experimental data made in Ref.~\cite{normann15}  converges around $\zeta_0 \sim 10^6~$ Pa s.

Let us make a conceptual remark on the hydrodynamic approach to the dark energy problem in general. In the literature one will often see that the represented by a self-interacting scalar field; the identification with some kind of "fluid" thereafter being made on a formal basis. One may ask: is the hydrodynamic picture merely a secondary step, following the more profound scalar field theory approach? In our opinion this is hardly so. All the time that the cosmic medium is modeled as a fluid in standard cosmology, the same hydrodynamic picture should be applicable also to its dark component. The hydrodynamic formalism has generally proved to be very robust.

In the next section we consider the future development of the cosmic dark fluid, assuming that the conditions are laminar. In Sect.~\ref{sec:3} we consider the thermodynamic formalism, deriving in Eq,~(\ref{23}) a differential equation determining the time-dependent temperature  $T(t)$.  Based upon the estimate (\ref{7}) for the bulk viscosity, we find that our universe can to a crude approximation  be categorized as a low-viscosity fluid, in view of  the condition (\ref{30}) for low viscosities. The evaluation of $T(t)$ becomes therewith facilitated. In Sect.~\ref{sec:4} we consider the transition to turbulence at $t=t_*$, and derive the positive temperature jump resulting if $\zeta_{\rm turb}>\zeta$. This is qualitatively as we would expect, and indicate that the theory is physically reasonable. Also, the   entropy production at the transition is considered. Finally we write the energy-momentum tensor for the fluid on a covariant form, and emphasize the formal relationship to a Maxwell viscoelastic fluid.

We will focus on the late universe, $t\geq 0$, although   occasionally we touch upon  the analytic continuation of formulas to the past universe, $t<0$.

\section{The laminar epoch} \label{sec:2}

As mentioned, we will assume that  $w$  is constant. Also, we take
\begin{equation}
\zeta =\zeta_0 \label{x}
\end{equation}
 to be a constant. This assumption is clearly attractive because of its mathematical simplicity. One may in addition ask: is the assumption reasonable from a physical viewpoint also? We recall from ordinary hydrodynamics that the kinematic viscosity $\nu=\eta/\rho$ associated with a shear viscosity $\eta$ is commonly taken to be proportional to the mean free path $l$, at a given temperature. If $l$ is assumed to grow with the scale factor $a$ in the expanding universe, then one would expect $\eta$, as well as its bulk companion $\zeta$, to grow also, in apparent conflict with Eq.~(\ref{x}). However, the behavior of the fluid is more complex, as is demonstrated clearly in the case of a phantom fluid: the density is increasing significantly in the late universe. It is difficult to envisage that such a behavior  should be compatible with a growing free path. The natural conclusion here is that the situation is too complex to apply the simple estimates made use of in common kinetic theory. The expression (\ref{x}) should not be abandoned on physical grounds. Actually,  the constant bulk viscosity model has quite frequently been made use of in the literature.

 Assume now that  $U^\mu$ is the four-velocity of the fluid; in comoving coordinates $U^0=1, U^i=0.$  We assume spatially flat FRW space, and put the cosmological constant $\Lambda$ equal to zero. Since the FRW space is homogeneous, there is no conduction of heat ($Q^\mu=0$); the rotation and shear tensors both vanish ($\omega_{\mu\nu}=\sigma_{\mu\nu}=0$), and the scalar expansion $\theta ={U^\mu}_{;\mu}=3H$ with $H=\dot{a}/a$ the Hubble parameter. The conservation equation for energy is ${T^{0\nu}}_{;\nu}=0$ with $T^{\mu\nu}$ the fluid's energy-momentum tensor. Thus  in comoving coordinates
 \begin{equation}
 \dot{\rho}+(\rho+p)\theta=\zeta \theta. \label{8}
 \end{equation}
 The continuity equation  is ${(nU^\mu)}_{;\mu}=0$, $n$ being the particle density,  meaning locally that
 \begin{equation}
 \dot{n}+n\theta=0. \label{9}
 \end{equation}
  Friedmann's equations are
 \begin{equation}
 3H^2=8\pi G\rho, \label{10}
 \end{equation}
 \begin{equation}
 \frac{2\ddot{a}}{a}+H^2=-8\pi G(p-\zeta \theta). \label{11}
 \end{equation}
Then taking into account the equation of state, we obtain the governing equation for the scalar expansion
\begin{equation}
\dot{\theta}(t)+\frac{1}{2}\alpha \theta^2(t)-12\pi G \zeta \theta(t)=0. \label{12}
\end{equation}
The solution for $H$ is
\begin{equation}
H=\frac{H_0e^{t/t_c}}{1+\frac{3}{2}\alpha H_0t_c(e^{t/t_c}-1)}, \label{13}
\end{equation}
where $t_c$ is the 'viscosity time',
\begin{equation}
t_c=\frac{1}{12\pi G\zeta}. \label{16}
\end{equation}
The expression (\ref{13}) deserves further attention. Although derived for the late universe, $t>0$, it should hold formally for $t<0$ also. At first sight one might expect it to yield $\dot{H}>0$, in conflict with observation, but this is not so as a more careful consideration shows: by calculating the logarithmic time derivative of $H$ near the present time $t=0$ one gets
\begin{equation}
\frac{\dot{H}}{H}=\frac{12\pi G\zeta}{c^2}-\frac{3}{2}\alpha H_0, \quad t \rightarrow 0.
\end{equation}
Choosing $\zeta_0 = 10^5~$Pa s as inferred from the analysis of Wang and Meng \cite{wang14}, and inserting $H_0=67.80$ km s$^{-1}$ Mpc$^{-1}=2.20\times 10^{-18}$ s$^{-1}$, one gets
\begin{equation}
\frac{\dot{H}}{H} \rightarrow 2.79(1-1200\alpha)\times 10^{-21}~\rm{s}^{-1}.
\end{equation}
That is, if $\alpha$ is positive and greater than about 0.001, then $\dot{H}<0$ in qualitative agreement  with experiment. If we choose $\zeta_0 = 10^6~$Pa s instead, the condition on $\alpha$ becomes more stringent, $\alpha  >0.01.$ In this way, Eq.~(\ref{13}) provides a useful test of the model versus experiment.

Let us now write down the solutions for the scale factor, and the density,
\begin{equation}
a=a_0\left[1+\frac{3}{2}\alpha H_0t_c(e^{t/t_c}-1)\right]^{\frac{2}{3\alpha}}, \label{14}
\end{equation}
\begin{equation}
\rho=\frac{\rho_0\,e^{2t/t_c}}{\left[1+\frac{3}{2}\alpha H_0t_c(e^{t/t_c}-1)\right]^2 }, \quad \label{15}
\end{equation}

From these equations it follows that if one admits $\alpha <0$, then $H\rightarrow \infty, a\rightarrow \infty$, and $\rho \rightarrow \infty$ at a finite value $t=t_s$, where
\begin{equation}
t_s=t_c\ln \left( 1+\frac{2}{3|\alpha| H_0 t_c}\right). \label{17}
\end{equation}
We thus have to do with a big rip. As mentioned above, we shall however assume that at some time $t=t_* <t_s$  there occurs a transition  into a turbulent state of motion. Before considering this point, we need some thermodynamic information. This is the subject of the next section.

\section{Thermodynamic considerations}
 \label{sec:3}
We start from the thermodynamic identity in comoving coordinates,
\begin{equation}
k_B\dot{\sigma}=\frac{1}{nT}\left[ \dot{\rho}-\frac{\rho+p}{n}\dot{n} \right], \label{18}
\end{equation}
where  $\sigma$ is the nondimensional entropy per particle. We take  $n$ and $T$ as independent coordinates, $\rho=\rho(n,T)$, and  exploit that $d\sigma$ is an exact differential \cite{weinberg71}. This implies that
\begin{equation}
\rho +p=n\left( \frac{\partial \rho}{\partial n}\right)_T+T\left( \frac{\partial p}{\partial T}\right)_n. \label{19}
\end{equation}
We can now calculate the relative time derivative $\dot{T}/T$, by starting from
\begin{equation}
\dot{T}=\frac{1}{(\partial \rho/\partial T)_n}\left[ \dot{\rho}-\left(\frac{\partial p}{\partial n}\right)_T\dot{n}\right]. \label{20}
\end{equation}
Inserting $\dot \rho$ from Eq.~(\ref{8}) and $\dot n$ from Eq.~(\ref{9}), and then using Eq.~(\ref{19}), we get
\begin{equation}
\frac{\dot T}{T}=-\frac{(\partial p/\partial T)_n\theta}{(\partial \rho/\partial T)_n }+\frac{\zeta \theta^2}{T(\partial \rho/\partial T)_n}, \label{21}
\end{equation}
which implies
\begin{equation}
\frac{\dot T}{T}= - \left( \frac{\partial p}{\partial \rho}\right)_n\theta+
\frac{\zeta \theta^2}{T(\partial \rho/\partial T)_n}. \label{22}
\end{equation}
So far, we have not involved the equation of state. Inserting $p=(-1+\alpha)\rho$ we obtain
\begin{equation}
\frac{d}{dt}\ln \left[\frac{T}{T_0}\left(1+\frac{3}{2}\alpha H_0t_c(e^{t/t_c}-1)\right)^{-\frac{2(1-\alpha)}{\alpha}}\right]= \frac{\zeta \theta^2}{T(\partial \rho/\partial T)_n}. \label{23}
\end{equation}
This equation determines the time dependence of $T$ for a dark fluid, when  the bulk viscosity $\zeta$ is constant. Whereas Eqs.~(\ref{21}) and (\ref{22}) are known in the literature, Eq.~(\ref{23}) is according to our knowledge new.

In the following we will assume that the viscosity is small, in the sense that it is sufficient to work to the first order in $\zeta$ on the right hand side of Eq.~(\ref{23}). We can then make use of the $\zeta=0$ equations,
\begin{equation}
H=\frac{H_0}{1+3\alpha H_0t}, \label{24}
\end{equation}
\begin{equation}
a=a_0\left( 1+3\alpha H_0 t\right)^{\frac{2}{3\alpha}}, \label{25}
\end{equation}
\begin{equation}
\rho=\frac{\rho_0}{(1+3\alpha H_0t)^2}, \label{26}
\end{equation}
from which  ($(\partial \rho/\partial T)_n \rightarrow d\rho/dT)$
\begin{equation}
\frac{d\rho}{dT} \rightarrow \frac{-\alpha}{1-\alpha}\frac{\rho}{T}. \label{27}
\end{equation}
It turns out that the right hand side of Eq.~(\ref{23}) becomes in this approximation a {\it constant},
\begin{equation}
\frac{\zeta \theta^2}{T(\partial \rho/\partial T)_n} \rightarrow -\frac{1-\alpha}{\alpha}\,\frac{\zeta \theta_0^2}{\rho_0}=-\frac{1-\alpha}{\alpha}(24\pi G\zeta).  \label{28}
\end{equation}
Then we can easily integrate Eq.~(\ref{23}) to get
\begin{equation}
T=
 T_0  \left[1+\frac{3}{2}\alpha H_0 t_c(e^{t/t_c}-1)\right]^{\frac{2(1-\alpha)}{\alpha}}   \exp \left(-\frac{1-\alpha}{\alpha}\frac{2t}{t_c}\right).
\label{29}
\end{equation}
We shall take the assumed smallness of the viscosity to imply that the following condition is satisfied:
\begin{equation}
 |\alpha| H_0t_c \gg 1. \label{30}
 \end{equation}
 As seen from Eq.~(\ref{17}), this means that to the lowest order
 \begin{equation}
 t_s = \frac{2}{3|\alpha| H_0}. \label{31}
 \end{equation}
The rip time  is accordingly much smaller than the viscosity time,
\begin{equation}
\frac{t_s}{t_c}=\frac{2}{3|\alpha| H_0t_c} \ll 1. \label{32}
\end{equation}
In turn, this implies that the argument in the last exponential in  Eq.~(\ref{29}) remains small in the whole time span $0<t<t_s$, except from cases where $\alpha$ is very close to zero.

It is of interest to check to what extent the condition  (\ref{30}) is satisfied in our universe. Using the estimate $\zeta = 10^5~$Pa s, we get for the viscosity time (in
dimensional units)
\begin{equation}
t_c=\frac{c^2}{12\pi G \zeta} = 3.58\times 10^{20}~\rm s. \label{33}
\end{equation}
Considering first $\alpha>0$ (quintessence region):  the maximum value $\alpha_{\rm max}$ of $\alpha$ can according to Eq.~(\ref{5}) be estimated as
\begin{equation}
\alpha_{\rm max} = -0.019+0.075=0.056. \label{34}
\end{equation}
Then, again  with $H_0 =2.20\times 10^{-18}$ s$^{-1}$, we  obtain $H_0t_c=787$ and
\begin{equation}
\alpha_{\rm max}H_0 t_c = 44. \label{35}
\end{equation}
 The condition (\ref{30}) is thus  roughly satisfied.

  On the other side of the phantom barrier, if $\alpha <0$ (phantom region), we see from Eq.~(\ref{5}) that $\alpha_{\rm min}=-0.019-0.080=-0.099$, giving
 \begin{equation}
 |\alpha_{\rm min}|H_0t_c = 78,
 \end{equation}
 so that the  condition (\ref{30}) is  somewhat better satisfied in this case.

 As shown in Ref.~\cite{brevik15} the estimated value for the viscosity, in conjunction with  a positive value of $\alpha$ (the first case discussed above),  is of the right order of magnitude  to drive the fluid through the phantom barrier into the phantom region even if it starts  from the quintessence region at $t=0$.

\section{Transition to the  turbulent epoch at $t=t_*$} \label{sec:4}
\subsection{Change in  temperature }

 Assume, as mentioned above, that the laminar flow starts with $\alpha<0$ at the initial time $t=0$ and transforms to a turbulent flow at the  instant $t_* <t_s$.  We assume the equation-of-state parameter to be a constant, called $w_{\rm turb}$, also in the turbulent region. For $t>t_*$ thus
\begin{equation}
p_{\rm turb}=w_{\rm turb}\,\rho_{\rm turb}, \label{36}
\end{equation}
 and in  analogy to Eq.~(\ref{6}), we define $\alpha_{\rm turb}$ according to
 \begin{equation}
 w_{\rm turb}=-1+\alpha_{\rm turb}. \label{37}
 \end{equation}
 These equations are introduced  from analogy reasons. Very little seems actually to be known about this transition to turbulence. As mentioned above, a fundamental theory for the onset of turbulence is lacking even in the case of ordinary turbulence. It may be instructive to recall how the eddy viscosity is introduced in the latter case: if $\nu_e$ denotes  the turbulent analogue of the kinematic viscosity $\nu=\eta/\rho$ in laminar theory, one writes
 \begin{equation}
 \nu_e=u' l_m, \label{y}
 \end{equation}
 where $u'$ is a typical value of the fluctuating velocity, and the length parameter $l_m$ called the Prandtl mixing length. The equation (\ref{y}) is important for the construction of the eddy viscosity and the Reynolds stresses . We think  it is physically natural to adopt this picture also for the dark fluid. The onset of instability in the violent motion of the dark fluid as the future singularity is approached, seems almost unavoidable if a fluid picture is to be maintained  in this epoch at all.

We see that Eq.~(\ref{23}) permits us to calculate the changes of physical quantities across $t=t_*$ in a convenient way. This feature provides the link to the Maxwell fluid theory shown in  Eq.~(\ref{1}), where the presence of the material parameter $\lambda$ allowed the  integration across the singularity to be done.

 One has to observe which physical quantities can change across the singularity. The total energy density  will not change upon a laminar-turbulent transition; nor does the scale factor or its time derivative change suddenly. Thus the set of variables $\{\rho, a, \theta\}$ is  continuous across $t_*$. What  can change, is the set $\{ p,T\}$, and of course the specific entropy $\sigma$.

We will restrict ourselves to the first order approximation in the viscosities $\zeta$ and $\zeta_{\rm turb}$. The expression  (\ref{28}) is  useful, as above, and  can be inserted on the right hand side of Eq.~(\ref{23}) (where  the equation-of-state parameter, and the viscosity, depend on time in the laminar/turbulent process). The right hand side of Eq.~(\ref{23}) changes abruptly during the transition. We may represent that mathematically as the time derivative of a unit step function taken at argument $(t-t_*)$. Upon integration across $t_*$ we  obtain the difference between the turbulent and laminar values taken respectively at $t_*^+$ and $t_*^-$. To avoid mathematical complexity, we shall   assume that the equation-of-state parameters are the same,
\begin{equation}
\alpha_{\rm turb}=\alpha. \label{38}
\end{equation}

 The integration across $t_*$ now yields
\begin{equation}
\frac{T_*^+}{T_*^-}=\exp\left( 24\pi G \,\frac{1+|\alpha|}{|\alpha|}
( \zeta_{\rm turb}-\zeta)\right) \label{39}
\end{equation}
(the small change in $t_c$ is negligible).

The result (\ref{39}) is actually quite natural physically. If $\zeta_{\rm turb}$ is greater than $\zeta$, what we should  expect, the cosmic fluid experiences a positive temperature jump, as is always the case for an  irreversible process in fluid mechanics. This result also provides a physical support for our assumption (\ref{38}) above.

\subsection{Change in entropy}

It is worthwhile to consider also the entropy change in the laminar/turbulent transition, making use of the standard formula for the local entropy rate of change for a single particle,
\begin{equation}
\dot{\sigma}=\frac{\theta^2}{k_Bn}\frac{\zeta}{T}. \label{42}
\end{equation}
As mentioned above, neither $\theta$ nor $n$ can change during the transition. Then using again the same argument as before, implying here  that $\zeta/T$ can be represented as the time derivative of a unit step function during the abrupt change, we find for the change $\Delta \sigma$ of specific entropy
\begin{equation}
\Delta \sigma=\frac{\theta^2(t_*)}{k_Bn(t_*)}\left(\frac{\zeta_{\rm turb}}{T_*^+}-\frac{\zeta}{T_*^-}\right). \label{43}
\end{equation}
 According to usual thermodynamics for an irreversible process, we must expect that   $\Delta \sigma$ is greater than zero, $ \zeta_{\rm turb}/T_*^+ > \zeta/T_*^-.$   That means, the jump in viscosity must dominate the jump in temperature across $t=t_*$.

\section{Summary} \label{sec:5}

Assuming zero shear viscosity and constant bulk viscosity, we have considered the evolution of the late universe ($t\ge 0$), adopting a one-component fluid model. The fluid was initially assumed laminar, with $\zeta$ the bulk viscosity. Experimental data show that the equation-of-state parameter $w$  lies quite close to $-1$, thus the magnitude of $\alpha =1+w$ is small. If initially $\alpha <0$ (phantom region) it has been known since the analysis of Caldwell et al. \cite{caldwell03} that the fluid is driven into a future singularity. We  assumed, however, that before this event takes place there occurs a transition into a turbulent state of motion at a definite time $t_*$, whereafter  the equation-of-state  parameter $w_{\rm turb}$ stays  constant. In such a scheme, it has been shown  earlier \cite{brevik12a} that the transition to turbulence can prevent the fluid from entering the future singularity at all.

In our final formulas we assumed for mathematical simplicity as well as for physical reasons that  $w_{\rm turb}=w$ (or $\alpha_{\rm turb}=\alpha$).  In Eq.~(\ref{39}) we derived a formula  for the temperature jump of the fluid in a typical laminar/turbulent fluid transition. The result appears to be physically natural, as it corresponds to a positive temperature jump when $\zeta_{\rm turb}>\zeta$, and an associated positive entropy production in the transition.

  As a by-product we derived an expression (\ref{23}) - to our knowledge not derived before -   for the time dependence of the temperature $T(t)$ in the laminar region. A simplifying factor in the evaluation of $T(t)$ is that, in view of recent experimental analysis of Wang and Meng \cite{wang14} and others, the cosmic fluid may roughly be approximated by a low viscosity fluid, thus permitting one to work to the first perturbative order in the viscosity.

Equation (\ref{23}) is  analogous to what is found for a viscoelastic Maxwell fluid,  as outlined in Sect.~\ref{sec:1}.  We think this property is worth emphasizing.  To our knowledge, this analogy has so far been left unnoticed in the literature.

Finally, let us write down a compact relativistic  expression for the fluid's energy-momentum tensor, valid for the whole region $0<t<\infty$:
\begin{equation}
T_{\mu\nu}=\rho U_\mu U_\nu+ph_{\mu\nu}-\zeta \theta [1-\Theta(t-t_*)]h_{\mu\nu}-\zeta_{\rm turb}\theta \Theta(t-t_*)h_{\mu\nu} \label{44}
\end{equation}
(recall that $\theta={U^\mu}_{;\mu}$ is the scalar expansion). Further, $h_{\mu\nu}=g_{\mu\nu}+U_\mu U_\nu$ is the projection tensor, and $\Theta(t)$ is the step function: $\Theta(t)=0$ if $t<0$ and $\Theta(t)=1$ if $t>0$. The expression (\ref{44}) is covariant, except that the time $t$ in comoving coordinates is used to distinguish between the laminar and turbulent regions.


\newpage

\begin{thebibliography}{99}

\bibitem{das02} P. S. Das, Indian J. Pure Appl. Math. {\bf 33}, 21 (2002).
\bibitem{planck} Planck Collaboration. Planck 2015 results. XIII. arXiv:1502.01589.
\bibitem{caldwell03} R. R. Caldwell, M. Kamionkowski, and N. N. Weinberg, Phys. Rev. Lett. {\bf 91}, 071301 (2003).
\bibitem{nojiri03} S. Nojiri and S. D. Odintsov, Phys. Lett. B {\bf 562}, 147 (2003).
\bibitem{nojiri04} S. Nojiri and  S. D. Odintsov,   Phys. Rev. D  {\bf 70},103522 (2004).
\bibitem{frampton11} P. H. Frampton, K. J. Ludwick, and  R. J. Scherrer, Phys. Rev. D {\bf 84}, 063003 (2011).
\bibitem{brevik11} I. Brevik, E. Elizalde, S. Nojiri, and S. D. Odintsov, Phys. Rev. D {\bf 84}, 103508 (2011).
\bibitem{brevik12} I. Brevik, R. Myrzakulov, S. Nojiri, and S. D. Odintsov, Phys. Rev. D {\bf 86}, 063007 (2012).
\bibitem{frampton12} P. H. Frampton, K. J. Ludwick, and R. J. Scherrer, Phys. Rev. D {\bf 85}, 083001 (2012).
\bibitem{wei12} H. Wei, L. F. Wang, and X. J. Guo, Phys. Rev. D {\bf 86}, 083003 (2012).
\bibitem{brevik13} I. Brevik, A. V. Timoshkin, Y. Rabochaya, and S. Zerbini, Astrophys. Space Sci. {\bf 347}, 203 (2013).
\bibitem{brevik12a} I. Brevik, R. Myrzakulov, S. Nojiri, and S. D. Odintsov, Phys. Rev. D {\bf 86}, 063007 (2012).
\bibitem{brevik11a}I. Brevik, O. Gorbunova, S. Nojiri, and S. D. Odintsov, Eur. Phys. J. C {\bf 71}, 1629 (2011).
\bibitem{bini13} D. Bini, A. Geralico, D. Gregoris, and S. Succi, Eur. Phys. J. C {\bf 73}, 2334 (2013).
\bibitem{wang14} J. Wang and X. Meng, Mod. Phys. Lett. A {\bf 29}, 1450009 (2014).
\bibitem{velten12} H. Velten and D. J. Schwarz, Phys. Rev. D {\bf 86}, 083501 (2012).
\bibitem{sasidharan15} A. Sasidharan and T. K. Mathew, arXiv:1511.05287.
\bibitem{brevik15} I. Brevik, Entropy {\bf 17}, 6318 (2015).
\bibitem{normann15} B. D. Normann and I. Brevik, arXiv:1601.04519.
\bibitem{weinberg71} S. Weinberg, Astrophys. J. {\bf 168}, 175 (1971).



\end{thebibliography}
\end{document}